# Theoretical terms and theoretical objects of contemporary cosmology as intellectual artefacts[1]

Anastasiia Lazutkina

Lomonosov Moscow State University

Contemporary cosmology contains a set of theoretical objects that may be considered intellectual constructs in one sense or another. The aim of this paper is to examine them as intellectual artefacts and to provide a motivation for this examination that stems from debates internal to the field of physical cosmology.

Examining theoretical terms as intellectual artefacts requires fitting them into a conceptual scheme that features such a class of objects. An influential example of such a scheme is Karl Popper's three world ontology, which builds on his scientific realism. World 1 is the mind-independent world studied by the natural sciences, whereas World 2 is the realm of subjective mental phenomena, which somehow emerges from World 1 but is not reducible to it. World 2 then produces World 3, where intellectual artefacts, such as scientific theories, are to be found. In contrast to the private and subjective World 2, the objects of World 3 have a public, "autonomous" existence apart from the mental states that populate World 2. For example, scientific theories have logical relations and may have unnoticed consequences and inconsistencies. (Pera 2006, 273-4)

It may be claimed that all objects belonging to World 3 can be regarded as intellectual artefacts, but in this paper the focus is on individual *terms* employed in physical cosmology, and the claim regarding their artefactual status is restricted: here, only those terms that are employed in *false* theories are claimed to be clearly artefactual. This, in turn, requires the assumption that some theories *are* false, even though terms employed in these false theories are still used in our language, especially in philosophical discourse.

[1] I would like to thank the following individuals for their contributions to this paper: John Antturi, Alexey Ilyin, Stacy McGaugh and Mordehai Milgrom. Naturally, none of the aforementioned individuals should be held responsible for the opinions asserted here or expected to agree with those opinions.

*Dark matter* is one of the latest potential intellectual artefacts from the field of cosmology, and a description and evaluation of its theoretical role is the main focus of this paper. For this purpose, another methodological assumption is utilized here, namely a classification of theoretical objects taken from Russian logician E. K. Voishvillo. (2014, 13) In this classification, theoretical objects are contrasted with objects of observation. Although both are featured in scientific theories, whether an object is theoretical or an observed one, in this specific sense, depends on the basis on which they are employed in a theory. Objects of observation are taken as given at the moment of constructing a theory, whereas theoretical objects are introduced to a pre-existent theory for the purpose explaining some phenomenon.

Among theoretical objects, a further non-exhaustive division may be introduced:

1) Abstract objects represent some of the properties of real[2] objects, or their relationships, actually turned into independent objects of thought – they are the results of isolating abstraction (e.g. the rotation of the Sun, the lifetime of a star, mass, force).

2) Idealized objects, which are created from observed objects by the logical operation of idealization: properties are either removed or added to them (e.g. absolute instability, absolute black body).

3) Ideal objects are the results of creative activity of thought, which usually play an instrumental role in the construction of a theory (e.g. complex numbers, a system of coordinates). Unlike an idealized object, an ideal object is not constructed from a real object to which the operation of idealization is performed. Rather, there is nothing in empirical domain to begin with that could be considered a "non-ideal" counterpart to an ideal object.

---

[2] In this terminology, real is equated with observed. This usage of "real" is neutral with regard to the meta-level discussion concerning scientific realism and anti-realism. Here the basis for referring to observed objects as real is that observations are assumed to be undisputed.

4) Hypothetical objects are objects that enter into thinking when constructing explanatory theories of empirical phenomena. When such objects are postulated, they are unobserved, but with the development of science can be independently detected, thus gaining the status of real objects, or being fictitious are discarded from the theories. In the latter case, what remains of them are the previously mentioned terms that are taken to be intellectual artefacts (those terms that feature in false theories), which are said to refer to empty objects (e.g. phlogiston and caloric).

Notably, the need for hypothetical objects may arise when there are discrepancies between theoretical predictions and empirical data. A hypothetical object can be employed to explain this discrepancy and save the theory from modification. Of course, modifying the theory *is* another way of explaining such a discrepancy, as shall be demonstrated.

In the history of science, and astronomy in particular, there are several prominent examples of theory choice, where the choice was between modifying a theory or affirming the existence of a hypothetical object.

The existence of five planets had been recognized by observers since antiquity: Mercury, Venus, Mars, Jupiter and Saturn.[3] As Baum and Sheehan note (1997, 45), early modern astronomers had a tendency to attempt to reduce this system of observed planets to calculation in order to determine the laws describing their motions, rather than to assume that there might be unobserved objects affecting their orbits. By the late 18th century, even the particularly

---

[3] The fact that the Sun was included among them in the geocentric model and that their relative positions to each other were partially incorrect is here omitted for the sake of simplicity.

tricky case of Jupiter and Saturn's seemingly irregular orbits had been solved by Laplace, confirming the sufficiency of Newton's theory of gravitation (the then unresolved case of Mercury's advancing perihelion shall be addressed presently). (ibid., 38)

It was thus a surprise to the astronomical community, when the orbit of the planet Uranus, discovered in 1781 by Hanover-born amateur astronomer William Herschel, could not be fitted with the theory. The reluctance to postulate a hypothetical object affecting Uranus's orbit meant that even modifying Newton's law of gravitation was on the table[4], although it remained an unattractive option, since it had so famously triumphed in the face of previous difficulties. Thus, it was considered a last resort by Urbain Jean-Joseph Le Verrier, who instead focused his efforts on constructing a hypothetical object that would explain the discrepancy. Fortunately for Newton's theory, a planet matching exactly the predicted properties of this hypothetical object was observed on September 23, 1846 by Johann Gottfried Galle at the Royal Observatory of Prussia. Neptune had been discovered, in what Levenson (2015, Ch. 3) describes as "the climax of what was almost immediately understood to be the popular triumph of Newtonian science."

However, the case of Mercury's advancing perihelion continued to tarnish the otherwise immaculate precision of Newton's theory throughout the 19th century. Again, the existence of a hypothetical object, the planet Vulcan[5], was postulated by Le Verrier to save the theory. The existence of such a planet had never been confirmed, however, despite observations dating back to the ancient Chaldeans and Chinese. (Baum and Sheehan 1997, 146). Sunspots had been detected by astronomers, of course, and sometimes they had been even mistaken

---

[4] According to Levenson (2015, Ch. 3) as well as Baum and Sheehan (1997, 80), Friedrich Wilhelm Bessel suggested in passing that the gravitational constant G, previously fixed at $6.674 * 10^{-11}$ N * m² / kg², might vary with distance.
[5] It is unclear at which point the name Vulcan was suggested for this intra-Mercurial object, although it first appears in published sources only after Lescarbault's 1859 observation of a body moving across the edge of the Sun's disc. (Baum and Sheehan 1997, 156-157)

for planets. Nonetheless, their movement across the Sun's disk was never as rapid as that of a planet's motion relative to an Earth observer, which is why the French amateur astronomer Lerscarbault was astonished to find an object moving rapidly across the edge of the Sun's disk on March 26, 1859 (ibid., 147-148). After he communicated the observation to Le Verrier, controversy ensued. Although Le Verrier himself admitted that the object observed could not be sufficiently massive to account for the perturbation it supposedly caused to Mercury's orbit, he remained a firm believer in the existence of Vulcan. (ibid. 157) Further predicted appearances of Vulcan were failures, and supposed observations were sporadic. (ibd. 162-168) The explanatory need for such a planet disappeared in any case, when Einstein's theory of general relativity was confirmed. A notable feature of the case of Vulcan is that the planet had been "observed" before being deemed a mere intellectual artefact. Therefore, at least for some astronomers, Vulcan was an empirical object, which is a healthy reminder that human fallibility is not restricted to speculative theorizing but also observation.

In a contemporary, unresolved case, namely that of dark matter, cosmologists of today face a situation similar to the one in which astronomers found themselves in the two previous examples. In the 1930s, the Swiss astronomer Fritz Zwicky studied the galaxy cluster Coma and found that the visible mass is insufficient to account for the observed gravitational effects. Galaxies in the cluster move at very high velocities with respect to their center of mass. When the mass required to keep the cluster together is calculated, the result is over one hundred times greater than the mass of the observed cluster. Zwicky suggested, that there must be invisible mass, *dunkle Materie* (dark matter) that provides the gravitational pull. The observations did not fit the theory, so a hypothetical object was introduced. (Zwicky 1938, 243-246)

Forty years later, Vera Rubin, studying individual galaxies, noticed that on the periphery of galaxies, the velocity at which stars and gas clouds rotate around the center of spiral galaxies is unexpectedly high. Based on Newton's laws, when moving from the galactic center to its periphery, the rotational velocity of the galactic objects is inversely proportional to the square root of the distance from the center. However, observations show that the rate does not fall, but rather it is constant. Thus, the rotation curve produced by Newtonian laws does not match the observational records. (Rubin and Ford 1970, 394)

Rubin's discovery was taken to be additional evidence for the existence of dark matter. Lately, many more arguments have been presented for its existence. The most recent evidence is based on the theory of general relativity and called the method of gravitational lensing. One consequence of general relativity is the assertion that any massive object warps space. If radiation is received from a distant galaxy, a galaxy cluster positioned between the source of the radiation and the observer bends the rays. Based on how much deformation occurs, it is possible to calculate the mass of the object distorting the radiation. Again, these calculations result in the conclusion that there must be more mass than that which we can detect.

These considerations have resulted in the wide acceptance of the so-called standard cosmological model, the $\Lambda$CDM (lambda-cold dark matter)[6] model, which predicts that 4.9% of the universe consists of baryons - of which all visible objects are made - and weakly interacting particles, neutrinos. The remaining 95.1% are hypothetical objects: 26.8% is dark matter, an object that cannot be directly detected, as it does not interact with electromagnetic radiation, but the existence of which is postulated on the basis of observable

---

[6] The Greek letter lambda refers to a cosmological constant, whereas "cold" refers to the types of hypothetical particles in this dark matter model.

gravitational effects. Finally, 68.3% of the universe consists of dark energy, introduced to explain the apparent expansion of the universe. (Ade et al. 2013)

It is not difficult to find prominent examples of dark matter being referred to as an object whose existence has already been confirmed, although no observations of it have ever been made despite decades of attempts. (Milgrom 2014) Several hypothetical particles have been proposed as candidates for filling the theoretical role of dark matter. Corresponding detectors have been built and continuous experiments have been run, but the results have been inconclusive – due to the very theoretically laden method required to observe something that does not interact with electromagnetic radiation, there are far too many alternative interpretations of the data that cannot be ruled out.

Perhaps the most promising alternative to dark matter has nothing to do with alternative interpretations of data gathered by dark matter detectors, however. Instead, the same option remains as in the case of Uranus and Mercury, i.e. modifying the theory whose predictions do not match the observations. Once again, the theory in question is the current theory of gravitation. Modified Newtonian dynamics (MOND) is often considered the strongest contender for solving the so-called mass discrepancy problem (or acceleration discrepancy, as some MOND proponents prefer to call it), because unlike theories postulating dark matter, it has actually managed to produce successful, specific, novel predictions, especially at the scale of individual galaxies. (Milgrom, 2014)

In 1983, MOND creator Mordechai Milgrom introduced a new parameter to the laws of Newton and produced novel theoretical predictions that turned out to be consistent with observations made several years later. Milgrom suggested that Newtonian dynamics is applicable only to large accelerations that can be found, for example, in our solar system. As soon a system with small accelerations is examined, there are discrepancies. A parameter introduced by Milgrom, critical acceleration, – $a_0$ – is a new constant ($a_0 = 1.2 \times 10^{-10}$ m / s$^2$). It separates small acceleration, which is less than $a_0$, and large acceleration, if it is greater than $a_0$. (ibid.) In accordance with this, a modification is made to Newton's laws:

> "When the acceleration is much larger than $a_0$, Newton's second law applies as usual: force is proportional to acceleration. But when the acceleration is small compared with $a_0$, Newton's second law is altered: force becomes proportional to the square of the acceleration. By this scheme, the force needed to impart a given acceleration is always smaller than Newtonian dynamics requires. To account for the observed accelerations in galaxies, MOND predicts a smaller force—hence, less gravity-producing mass—than Newtonian dynamics does … In this way, it can eliminate the need for dark matter." (Milgrom 2002, 46)

As can be seen, the rivalry between MOND and dark matter theories has similarities to the famous cases in the history of astronomy, where choices were made regarding the introduction of novel intellectual artifacts into the universe of real objects. An important difference is that in the historical examples, Newton's theory was able to overcome many apparent difficulties by producing novel predictions that aligned with later observational data to an exceptional degree, with Neptune's discovery standing out as an eminent example. Dark matter theories, in contrast, make no unique predictions, and thus lack a theoretical virtue crucial to the celebrated speculative constructs of the past. MOND, on the other hand, boasts several predictive successes, although its extension to the domain of relativistic and cosmological phenomena remains an incomplete task. (Milgrom 2014)

However, it remains a sociological fact that dark matter theories are dominant in the discipline, which naturally enough has turned some proponents of MOND to theorize on the sociological aspects of cosmology. (See, e.g. Lopez-Corredoira 2008) The situation can be summarized in the following way: for any given observation, if the observation is consistent with dark matter theories, dark matter theories are confirmed. If the observation is inconsistent with dark matter theories, then the observation must be wrong. If the observation is consistent with MOND, the observation must also be wrong. If the observation is inconsistent with MOND, MOND is falsified.[7]

In the face of such controversies, those with empiricist inclinations might want to suspend judgment, until decisive observational evidence is produced to rule out at least one of the competing theories. However, it is worth mentioning that here the relevant philosophical input might come not just from studying the history of science, but also from novel methodological reflection, such as applying certain tools developed by 20th century logicians. This would involve adding normative suggestions to the descriptive examination of theoretical terms as intellectual artefacts, which this paper has been concerned with.

In fact, the need for such methodological reform may be best understood by those keeping an eye on the historical development of the discipline, since they may be justifiably concerned by contemporary physicists' relative isolation from philosophy, when compared to those physicists who produced the most significant breakthroughs of the early 20th century, such as Mach, Einstein, Schrödinger, Heisenberg and others, who were keen practitioners of methodological reflection.

Bibliography:

---

[7] Stacy McGaugh (2012) has provided a flowchart that describes this dynamic. It is available at http://astroweb.case.edu/ssm/mond/LCDMMONDflowchart.png